\documentclass[prl,amsmath,amsfonts,amssymb,twocolumn,superscriptaddress,showpacs]{revtex4}

\usepackage{graphicx,psfrag}
\usepackage{dcolumn}
\usepackage{bm}
\usepackage{mathbbol}
\usepackage{makeidx}               
\usepackage{graphicx}              
\usepackage[bookmarksnumbered,colorlinks,plainpages,citecolor=blue,urlcolor=blue]{hyperref}
\usepackage{subfigure}
\usepackage{caption2}
\usepackage{latexsym}
\usepackage{amsmath}
\usepackage{amsfonts}
\usepackage{amssymb}
\usepackage{enumerate}
\usepackage{draftwatermark}

\newcommand{\uck}[1]{\o}

\newcommand{\ket}[1]{\mbox{$|#1\protect\rangle$}}

\newcommand{\bra}[1]{\mbox{$\protect\langle#1|$}}

\newcommand{\abs}[1]{\lvert#1\rvert}

\def\beq{\begin{equation}}
\def\eeq{\end{equation}}
\def\bea{\begin{eqnarray}}
\def\eea{\end{eqnarray}}

\begin{document}

\SetWatermarkText{Draft}
\SetWatermarkScale{8}
\SetWatermarkHorCenter{300pt}
\SetWatermarkVerCenter{400pt}

\begin{titlepage}
\date{\today}

\title{Tsallis entropy is natural in the formulation of quantum noise}

\author{Boaz Tamir \footnote{email: canjlm@actcom.co.il}}
\address{Faculty of interdisciplinary studies, Bar-Ilan University, Ramat Gan, Israel}
\address{ IYAR, Israel institute for advanced research, Rehovot, Israel}

\maketitle

\section*{Abstract}

In this paper we introduce an easy to compute upper bound on the Tsallis entropy of a density matrix describing a system coupled to a noise source. This suggests that the Tsallis entropy is most natural in the context of quantum information theory. Similarly we re-define 'entropy exchange' and use the same bound.

\end{titlepage}

\section{Tsallis entropy in quantum noise theory}

Let $\rho$ be a density matrix on the space $V$. Define the Tsallis entropy of index 2 \cite{Tsallis} to be:
\begin{equation}
 h(\rho) = 1-tr(\rho^2) 
\end{equation}

\noindent also known as the \emph{logical entropy} \cite{Ellerman1} \cite{Ellerman2} (see the appendix for the motivation of this definition). Clearly if $\rho$ is a pure state then $tr(\rho) = tr(\rho)^2=1$ and therefore $h(\rho)=0$.

Consider now the context of Operator Sum Representation for quatum noise (see \cite{Nielsen}). We start with the density $\rho=\rho(\ket{\psi})$ where $\ket{\psi}$ is pure, and an environment $E$. Suppose $\{ \ket{i} \}_i$ is an orthogonal basis of the environment. We also assume the environment is initially in the pure state $\ket{0}\bra{0}$. Next we operate by a unitary operator $\hat{U}$ which couples the system with its environment, hence introducing the noise:
\begin{equation}
\rho_E^{\hat{U}}(\ket{\psi}) = \hat{U} \left( \rho \otimes \ket{0}\bra{0} \right) \hat{U}^\dag 
\end{equation}
\noindent  Define:

\[ E_i = \bra{i}\hat{U} \ket{0}\]

\noindent Let:

\[ B_{i,j} = E_i \rho_E^{\hat{U}}E_j^\dag \]

\noindent  If we now trace out the environment we will get:
\begin{equation}
\tilde{\rho}^{\hat{U}} = \sum E_i \rho_E^{\hat{U}} E_i^\dag = \sum B_{i,i}
\end{equation}
The following theorem describes a measurement of the amount of noise introduced into the system by the environment:

\textbf{Theorem:} The logical entropy of the density matrix  $\tilde{\rho}^{\hat{U}} = \sum E_i \rho_E^{\hat{U}} E_i^\dag$  is bounded above by:

\begin{equation}
 h(\tilde{\rho}^{\hat{U}}) \leq \sum_{i\neq j} tr( B_{i,j} B_{i,j}^\dag)
\end{equation}

\noindent The above bound is the sum of the absolute squares of all elements which are off (below or above) the main blocks $ B_{i,i}$ of $\rho_E^{\hat{U}}(\ket{\psi})$.  Note that these off-main-blocks elements are related to the noise, if there is no coupling of the system with the environment these are all 0. Such a simple formula is a result of using the logical entropy, and does not exists for the von-Neumann entropy. This result stem out of a similar work by Ellerman \cite{Ellerman1}, and is given here a larger context. We will first show an example and then prove the theorem.

\textbf{Example: Amplitude damping}

We assume the environment is a two dimensional space.  Define the unitary operator to be \cite{Nielsen}:

\[ \hat{U} = \left( \begin{array}{cccc}1&0&0&0\\ 0&\cos(\theta)&0&-\sin(\theta)\\0&\sin(\theta)&0&\cos(\theta)\\0&0&1&0\end{array} \right) \]

\noindent We write $\rho$ as

\[ \rho = \left( \begin{array}{cc}a&b\\ \overline{b}&c\end{array} \right) \]

\noindent where $a$ and $c$ are real and positive, and $\overline{b}$ is the complex conjugate of $b$. Then:

\[  \hat{U}\left( \rho \otimes \ket{0} \bra{0} \right) \hat{U}^\dag =\]

\[=  \left( \begin{array}{cccc} a&b\cos(\theta)&b\sin(\theta)&0\\ \cos(\theta) \overline{b}&\cos^2(\theta) c&\cos(\theta) \sin(\theta) c&0\\\sin(\theta)\overline{b}&\sin(\theta)\cos(\theta) c &\sin^2(\theta) c &0\\0&0&0&0\end{array} \right) \]

\noindent If we now trace out the environment we will get two density matrices that corresponds to two operators $\hat{E}_0$ and $\hat{E}_1$, acting on $\rho$: 

\[ \hat{E}_0  \left( \begin{array}{cc}a&b\\ \overline{b}&c\end{array} \right) \hat{E}_0^\dag = \left( \begin{array}{cc}a&b \cos(\theta)\\ \cos(\theta)\overline{b}&\cos^2(\theta)c\end{array} \right)\] 

\[ \hat{E}_1  \left( \begin{array}{cc}a&b\\ \overline{b}&c\end{array} \right) \hat{E}_1^\dag = \left( \begin{array}{cc} \sin^2(\theta) c&0 \\ 0&0\end{array} \right)\]

\noindent where 

\[ \hat{E}_0 =  \left( \begin{array}{cc}1&0\\ 0&\cos(\theta)\end{array} \right) \]

\[ \hat{E}_1 =  \left( \begin{array}{cc}0&\sin(\theta)\\ 0&0\end{array} \right) \]

\noindent  The density $\tilde{\rho}$ following this noisy channel is:

\[\tilde{\rho} = \hat{E}_0  \left( \begin{array}{cc}a&b\\ \overline{b}&c\end{array} \right) \hat{E}_0^\dag  + \hat{E}_1  \left( \begin{array}{cc}a&b\\ \overline{b}&c\end{array} \right) \hat{E}_1^\dag \]

The theorem states that one can read an upper bound on the (logical) entropy of $\tilde{\rho}$  by simply adding twice the absolute squares of the off- 2 by2 block of $ \hat{U}\left( \rho \otimes \ket{0} \bra{0} \right) \hat{U}^\dag$, and the bound here is:  
\begin{equation}
 2 \abs{b}^2\sin^2(\theta) + 2 \cos^2(\theta)\sin^2(\theta) c^2 
\end{equation}
We can directly compute the trace of  $\tilde{\rho}^2$: 
\begin{equation}
 tr \tilde{\rho}^2 = a^2 + \sin^4(\theta) c^2 + 2 a c \sin^2(\theta) + 2 \abs{b}^2 \cos^2(\theta)+ \cos^4(\theta)c^2 
\end{equation}
\noindent and indeed we can verify that:
\begin{equation}
 h(\tilde{\rho}) = 1-tr \tilde{\rho}^2 \leq 2\abs{b}^2\sin^2(\theta) + 2 \cos^2(\theta)\sin^2(\theta) c^2 
\end{equation}

\textbf{Proof of the theorem:}  Observe that if we use a projective measurement on $\rho_E^{\hat{U}}(\ket{\psi})$, projecting the pure state into the blocks $B_i= V\otimes \ket{i}\bra{i}$ defined by the noise we will get:
\[ \sum \hat{P}_{B_i} \rho_E^{\hat{U}}(\ket{\psi}) \hat{P}_{B_i},\]

\noindent if we alternatively trace out the environment from $\rho_E^{\hat{U}}(\ket{\psi})$ we will get:

\[ \sum E_i \rho_E^{\hat{U}}(\ket{\psi}) E_i^\dag \]

\noindent The summand in both cases give the same set of density matrices, however in the case of projective measurement these density matrices have orthogonal support, whereas is the partial trace case they work on the same space. The theorem now is proved in two steps, first we show that this upper bound is exactly the logical entropy of the post projected pure state:

\begin{equation}
h\left( \sum \hat{P}_{B_i} \rho_E^{\hat{U}}(\ket{\psi}) \hat{P}_{B_i} \right) =  \sum_{i\neq j} tr( B_{i,j} B_{i,j}^\dag)
\end{equation}

\noindent  Second we will show that in general if a density matrix $\rho$ is a sum of density matrices then the logical entropy of ${\rho}$ is \emph{maxima}l if all components have orthogonal support and therefore: 
\begin{equation}
h\left( \sum E_i \rho_E^{\hat{U}}(\ket{\psi}) E_i^\dag \right) \leq h\left( \sum \hat{P}_{B_i} \rho_E^{\hat{U}}(\ket{\psi}) \hat{P}_{B_i} \right) 
\end{equation}


\textbf{Proposition 1:} Let $\rho$ be a density matrix, let the set of vectors $\{u_i\}$ be an orthogonal basis. Let  $\hat{\rho} = \sum \hat{P}_i \rho \hat{P}_i$ be a projective measurement on $\rho$. Let $B_i$ the partition blocks defined by $\hat{P}_i$ on $\{u_i\}$, then:
\begin{equation}
 tr \rho^2 = tr \left( \sum_i \hat{P}_i \rho \hat{P}_i \right)^2 + \sum_{x_{i,j}\not\in B_i} x_{i,j} x_{j,i}
\end{equation}
\noindent Observe that the elements in the last summand are complex conjugate since each is a reflection of the other with respect to the diagonal, and the matrix is a density operator, also since we can interchange i and j each element is counted twice. Having that in mind it is easy to prove the proposition by direct computation. 

\textbf{Corollary 1:} The logical entropy of a density matrix $\rho$  relate to the logical entropy of the post measurement density matrix $\hat{\rho}$ as follows: 
\begin{equation}
 h(\rho) = h\left(\sum_i \hat{P}_i \rho \hat{P}_i \right) -  \sum_{x_{i,j}\not\in B_i} x_{i,j} x_{j,i}
\end{equation}
\noindent In particular if $\rho$ is pure then the entropy of the post measurement density matrix $\hat{\rho}$ is twice the sum of squares of the off-block elements. This corollary was also discussed by Ellerman in \cite{Ellerman1}, there it was proved somewhat differently. Here we have implicitly used this result. If we apply a projective measurement on the space  $\hat{U} \left( \rho \otimes \ket{0}\bra{0} \right) \hat{U}^\dag$  projecting into the blocks $B_i$, then we get:

\[  \sum_i \hat{P}_i \hat{U} \left( \rho \otimes \ket{0}\bra{0} \right) \hat{U}^\dag\hat{P}_i \]

\noindent  In the example above if we project the pure state into the two blocks $B_i$ we get the two density matrices:

\[  \left( \begin{array}{cc}a&b \cos(\theta)\\ \cos(\theta)\overline{b}&\cos^2(\theta)c\end{array} \right)\] 

\[  \left( \begin{array}{cc} \sin^2(\theta) c&0 \\ 0&0\end{array} \right),\]  

\noindent  these two density matrices are now orthogonal! If we compute the logical entropies of the two components of  $\hat{\rho}= \hat{\rho}_1+\hat{\rho}_2$ (we use the notation $\tilde{\rho}$ to denote the density following the tracing out the environment, and $\hat{\rho}$ to denote the density following a projective measurement):

\[ tr(\hat{\rho}^2_1) = a^2 +2 \abs{b}^2\cos^2(\theta) + \cos^4(\theta) c^2 \]

\[ tr(\hat{\rho}^2_2) = \sin^4(\theta) c^2 \]

\noindent Twice the squares of the 2 by 2 off-block elements is:

\[ 2\abs{b}^2\sin^2(\theta) + 2 \cos^2(\theta)\sin^2(\theta) c^2 \]

\noindent and indeed it is easy to verify that in the orthogonal case:
\begin{equation}
 \hat{\rho} = 1-(tr(\hat{\rho}^2_1)+ tr(\hat{\rho}^2_2)) = 2\abs{b}^2\sin^2(\theta) + 2 \cos^2(\theta)\sin^2(\theta) c^2 
\end{equation}

\textbf{Corollary 2:} For any density matrix:
\begin{equation}
 h(\rho) \leq h\left(\sum_i \hat{P}_i \rho \hat{P}_i \right) 
\end{equation}
\noindent We will need this corollary in the proof of the next theorem. The entropy that we get by projecting the pure state into the blocks naturally produced by the noise interaction is the maximal one, while tracing out the environment will give a lower entropy. This will be clear following the next theorem.

\textbf{Proposition 2:} Let $\rho= p_1 \rho_1+ p_2 \rho_2$ be a density matrix where  $\rho_1$ and $\rho_2$ are two density matrices, then
\begin{equation}
h(\rho) \leq h(p_1,p_2) + p_1^2 h(\rho_1)+ p_2^2 h(\rho_2) 
\end{equation}
\noindent where the equality holds for $\rho_1$ and $\rho_2$ supported on orthogonal spaces, and $h(p_1,p_2)$ is the logical entropy of the distribution $(p_1,p_2)$:

\[ h(p_1,p_2) = 1-(p_1^2 + p_2^2) \] 

\textbf{Proof:}  (a) First assume that $\rho_1$ and $\rho_2$ have orthogonal support. Then

\[ \rho^2 = p_1^2 \rho_1^2 + p_2^2 \rho_2^2 \] 

\noindent Then taking the traces of both sides:

\[ 1-tr\rho^2 = 1- (p_1^2 tr \rho_1^2 + p_2^2  tr\rho_2^2 ) \]

\noindent Then:
\begin{equation}
 1-tr\rho^2 = 1- (p_1^2 +p_2^2) + p_1^2 (1-tr\rho_1^2)+ p_2^2  (1-tr\rho_2^2) =
\end{equation}
\[ = h(p_1,p_2) + p_1^2 h(\rho_1)+ p_2^2 h(\rho_2) \] 

\noindent which concludes this case.\\

\noindent (b) Assume $\rho_1$ and $\rho_2$ are pure, $\rho_i = \ket{\psi_i} \bra{\psi_i}$, then our system $S$ is described by:

\[ \rho = \sum p_i \ket{\psi_i} \bra{\psi_i}\]

\noindent We now purify $\rho$ by introducing another system B, with orthogonal states $\ket{i}$:

\[ \ket{SB} = \sum\sqrt{p_i} \ket{\psi_i} \ket{i} \]

\noindent Then we can write $\rho_{SB}= \ket{SB}\bra{SB}$.  Now, since $ \ket{SB}$ is pure we can use the Schmidt decomposition to get:

\[ h(\rho_S) = h(\rho_B) \]

\noindent where $\rho = \rho_S= tr_B \rho_{SB}$ and $\rho_B = tr_S \rho_{SB}$. Looking at  $\rho_B$ we get:
\begin{equation}
\rho_B = \sum_{i,j} \sum_l \left( \bra{l} \psi_i \rangle \bra{\psi_j} l\rangle \right) \ket{i}\bra{j} \sqrt{p_i} \sqrt{p_j}
\end{equation}
\noindent where $\{ \ket{l} \}$ is an orthognal basis for $S$. If we now measue $\rho_B$ in the $\ket{i}$ basis using the projections $\hat{P}_i = \ket{i}\bra{i}$ we get:
\begin{equation}
\hat{\rho}_B = \sum \hat{P}_i \rho_B \hat{P}_i = \sum p_i \ket{i}\bra{i}
\end{equation}
\noindent As we saw above (corollary 2), the logical entropy only increases following a projective measurement and therefore:
\begin{equation}
 h(\hat{\rho}_B) \geq h({\rho}_B)=  h(\rho)
\end{equation}
\noindent Using the above explicit formulation of  $\hat{\rho}_B$ we can deduce that $h(\hat{\rho}_B)= h(p_1,p_2)$. Now since $\ket{\psi_i}$ are both pure $h(\rho_i) =0$ and we can conclude this case.

\noindent (c) For the general case we can write:

\[ \rho = \sum p_i \rho_i \]

\noindent where

\[ \rho_i = \sum\lambda_{i,j} \ket{\lambda_{i,j}} \bra{\lambda_{i,j}} \]

\noindent where $\ket{\lambda}_{i,j}$ are orthogonal in the space of $\rho_i$. So $\rho$ can be written as a sum of pure states not necessarily all orthogonal:

\[ \rho = \sum p_i \lambda_{i,j} \ket{\lambda_{i,j}} \bra{\lambda_{i,j}}\]

Using the above case of pure states we can write:
\begin{equation}
h(\rho) \leq h(\{p_i \lambda_{i,j}\}_{i,j}) = 1-\sum_{i,j} p_i^2 \lambda_{i,j}^2 =
\end{equation}
 
\[ = (1-\sum_{i} p_i^2) + \sum_i p_i^2 \left(1-\sum_j \lambda_{i,j}^2\right) \] 

\noindent  which concludes the claim, and therefore the proof of the theorem. $\blacksquare$

\noindent We can now use the above theorem to define 'entropy exchange'. Let $\rho$ be a density matrix on our system $S$. We can purify $\rho$ by introducing an auxiliary system $R$, such that $\ket{RS}$ is pure, no let:

\[ \rho = \ket{RS}\bra{RS}   \]

\noindent Given an operator $\mathcal{E}$ on $\rho(\ket{RS})$ we can mock up its operation by introducing an environment $E$, initially in $\ket{0}\bra{0}$ eigenstate, as above. Define the exchange  entropy to be:

\[ h\left( tr_E ( \rho_E^{\hat{U}} (\ket{RS}))\right) \]

\noindent By the above theorem the exchange entropy is bounded above by:

\[  h( tr_E ( \rho_E^{\hat{U}} (\ket{RS}))) \leq \sum_{i\neq j} tr B_{i,j} B_{i,j}^\dag \]

\noindent where 

\[ B_{i,j} = E_i \rho_E^{\hat{U}} E_J^\dag \]

\noindent and 

\[ E_i = \bra{i} \hat{U} \ket{0} \]

\noindent The exchange entropy is important for the definitions of other quantum information quantities like \emph{coherent information}, and there is reason to believe that such quantities could also be delt easily with logical entropy.

\section{Appendix}

We will motivate here the definitions and the use of logical entropy in information theory in general and in quantum information theory in particular, we follow ideas presented in \cite{Ellerman}.\\

Let $P$ be a distribution on a set $\{u_1,...,u_n\}$ where $P(u_i) = p_i$.  Define the logical entropy $h(P)$ to be:
\begin{equation}
 h(P) = 1-\sum p_i^2 = \sum_{j\neq i} p_i p_j 
\end{equation}
\noindent Thus the logical entropy of a distribution is the sum of all distinctions, i.e. products of different pairs. It is the probability to get different results $(u_i,u_j)$ if we sample the distribution twice, these are also called dits and we also use the term dit(P).

 Let $\cup_i B_i = U$ be a partition $\pi_B$ of $U$, where $B_i\subseteq U$. We can similarly define the logical entropy of the partition $\pi_B$ to be: 
\begin{equation}
 h(\pi_B) = \sum_{u_k,u_j \in B_k x B_j, k\neq j} p_k p_j = dit(\pi_B) 
\end{equation}
\noindent Then
\begin{equation}
 h(\pi_B)= 1-  \sum_{u_k,u_j \in B_k x B_k} p_k p_j 
\end{equation}
There is a natural duality between set theory and partition theory. This is the main motivation in introducing the logical entropy (and its name). One can sense this duality looking at unions and intersections of dits. If $\pi_1$ and $\pi_2$ are two partitions then $dit(\pi_1) \cup  dit(\pi_2)$  are the distinct pairs that are in $\pi_1$ or in $\pi_2$ if a pair is not distinct in one partition, therefore in the same block there, it could still be distinct with respect to the other partition, therefore the union of two distinctions is a fine graining of both, which corresponds to an intersection in set theory. Similarly, the intersection of two dits,  $dit(\pi_1)\bigcap dit(\pi_2)$ are the distinct pair in both partitions and therefore corresponds to course graining or union in set theory. Hence the duality between partitions and sets with respect to the notion of distinctions. \\

We can extend the above definitions to the case of density matrices. The logical entropy of the density $\rho$ will be:

\[ h(\rho) = 1-tr(\rho^2) \]

Suppose now the vector space $V$ is spanned by the basis $\{u_1,...,u_n\}$.  We can use a partition $\{B_i\}$ on the set $U= \{u_1,...,u_n\}$ to define a projective measurement $\hat{P}=\sum\hat{P}_i $, such that $\hat{P}_i(V) = V_i$  and  $V_i$ is spanned by $u_i \in B_i$. 



\noindent  Let $\rho$ be a density matrix on $V$:

\[ \rho = \sum p_k \ket{u_k} \bra{u_k} \]

\noindent  Let $\hat{\rho}_{\hat{P}} = \sum \hat{P}_i \rho \hat{P}_i$  be the post measurement state for $\rho$ with respect to the projective measurement $\hat{P}$ defined above, then \cite{Ellerman1} \cite{Ellerman2}:

\begin{equation}
 h(\hat{\rho}_{\hat{P}}) = h(\pi_B)
\end{equation}

\section{Discussion}

We used the Operator Sum Representation of quantum noise to introduce an extremely easy to compute bound on the entropy of the density matrix following its interaction with the environment. Note that prior to the coupling with the environment the entropy of the density matrix is 0 (being pure).  Following the unitary interaction with the environment and just before tracing out the environment, the density matrix contains the information of the effects of the noise at its off block elements. Here we show that the right way to peal off this information is by using the Tsallis entropy (of index 2), or the logical entropy, reintroduced by Ellerman with its new motivation in partition theory. All this suggests a new definition for \emph{exchange entropy}, \emph{coherent information}, etc. \\

The simplicity gained by the use of Tsallis entropy is non accidental. It was recently shown by Ellerman \cite{Ellerman1} that this entropy formulation (which he coined logical entropy) is most natural in quantum measurement theory. It is deeply rooted in the duality between set theory and partition theory.  Projective measurements define partitions, and therefore Tsallis entropy on partitions could be generalized to quantum measurement theory.

\end{document}